\documentclass[]{aa}
\usepackage{epsfig}
\usepackage{natbib}
\usepackage{graphicx}

\begin{document}

\title{The INTEGRAL/IBIS Scientific Data Analysis\thanks{Based on
observations with INTEGRAL, an ESA project with instruments and science
data centre funded by ESA member states (especially the PI countries:
Denmark, France, Germany, Italy, Switzerland, Spain), Czech Republic and
Poland, and with the participation of Russia and the USA.}}

   \author{
A. Goldwurm \inst{1}, P. David \inst{1},   
L. Foschini \inst{2},  A. Gros \inst{1}, 
P. Laurent \inst{1}, A. Sauvageon \inst{1}, 
A.J. Bird \inst{3}, 
L. Lerusse \inst{4}, N. Produit \inst{4}
}

   \offprints{A. Goldwurm : agoldwurm@cea.fr}

   \institute{
   CEA Saclay, DSM/DAPNIA/SAp, F-91191 Gif sur Yvette Cedex, France
   \and IASF/CNR, sezione di Bologna, via Gobetti 101, 40129 Bologna, Italy 
   \and School of Physics and Astronomy, University of Southampton, Highfield,
SO17 1BJ, UK
   \and Integral Science Data Center, Chemin d'Ecogia, 16, CH-1290 Versoix, Switzerland
   }

\date{Received ; accepted}

\authorrunning{Goldwurm et al.}
\titlerunning{IBIS Science Data Analysis }

\abstract{
The gamma-ray astronomical observatory INTEGRAL, succesfully launched
on 17$^{\rm th}$ October 2002, carries two large gamma-ray telescopes. 
One of them is the coded-mask imaging 
gamma-ray telescope onboard the INTEGRAL satellite (IBIS) which provides
high-resolution ($\approx$~12$'$) sky images of 29$^{\circ}$~$\times$~29$^{\circ}$
in the energy range from 15 keV to 10 MeV with typical on-axis sensitivity 
of $\approx$ 1~mCrab at 100 keV (3$\sigma$, 10$^6$~s exposure).
We report here the general description of the IBIS coded-mask imaging system 
and of the standard IBIS science data analysis procedures. 
These procedures reconstruct, clean and combine IBIS sky images 
providing at the same time detection, identification and preliminary analysis of point-like 
sources present in the field. Spectral extraction has also been implemented and is based on 
simultaneous fitting of source and background shadowgram models to detector images.
The procedures are illustrated using some of the IBIS data 
collected during the inflight calibrations and present performance is discussed.
The analysis programs described here have been integrated as instrument specific 
software in the Integral Science Data Center (ISDC) analysis software packages 
currently used for the Quick Look, Standard and Off-line Scientific Analysis.
}
\maketitle


\section{The IBIS coded aperture imaging system}
The IBIS telescope (Imager on Board of the INTEGRAL Satellite) (Ubertini et al. 2003), 
launched onboard the ESA gamma-ray space mission
INTEGRAL (Winkler et al. 2003) on October 2002, is a hard-X ray/soft $\gamma$-ray telescope 
based on a coded aperture imaging system (Goldwurm et al. 2001).

In coded aperture telescopes (Fig. 1) (Caroli et al. 1987, Goldwurm 1995, Skinner 1995)
the source radiation is spatially modulated 
by a mask of opaque and transparent elements before 
being recorded by a position sensitive detector,
allowing simultaneous measurement of source plus background (detector area corresponding to 
the mask holes) and background fluxes (detector corresponding to the opaque elements).
Mask patterns are designed to allow each source in the field of view (FOV) 
to cast a unique shadowgram on the detector, in order to avoid ambiguities in the reconstruction
of the sky image. This reconstruction (deconvolution)
is generally based on a correlation procedure between the recorded image
and a decoding array derived from the mask pattern.
Assuming a perfect position sensitive detector plane (infinite spatial resolution), 
the angular resolution of such a system is then defined by the angle subtended
by one hole at the detector. The sensitive area instead depends on
the number of all transparent elements of the mask viewed by the detector.
In the gamma-ray domain where the count rate is dominated by the background,
the optimum transparent fraction is one half.
The field of view (sky region where source radiation is modulated by the mask)
is determined by the mask and the detector dimensions and their respective distance.
To optimize the sensitive area of the detector and have large FOVs, 
masks larger than the detector plane are usually employed. 
The FOV is thus divided in two parts. The fully coded (FC) FOV for which all 
source radiation directed towards the detector plane 
is modulated by the mask and the Partially Coded (PC)
FOV for which only a fraction of it is modulated by the mask. 
The rest, if detected, cannot be easily distinguished from the background. 
If holes are uniformly distributed the sensitivity 
is approximately constant in the FCFOV and decreases in the PCFOV.

Representing the mask with an array $M$ of 1 (transparent) and 0
(opaque) elements, the detector array $D$ is given by the convolution 
of the sky image $S$ with $M$ plus an unmodulated background array term $B$: 
~$D = S \star M + B$.
If $M$ has a
{\it correlation inverse} array $G$ such that $M \star G = \delta$-function, 
then we can reconstruct the sky by performing the following simple operation
$$S' = D \star G = S \star M \star G + B \star G = S \star \delta + B \star G = S + B \star G$$
and $S'$ differs from $S$ only by the $B \star G$ term.
In the case the total mask $M$ is derived from a cyclic
replication of the same basic pattern and the background is given by a 
flat array $B$, the term $B \star G$ is a constant and can be removed.
Mask patterns with such properties, including the {\it uniformly redundant 
arrays} (URA), were found in the 70s (Fenimore \& Cannon 1978) 
and then succesfully employed in X/$\gamma$-ray telescopes onboard 
several high energy missions (Caroli et al. 1987, Paul et al. 1991).

\begin{figure}
\centering
\centerline{\epsfig{figure=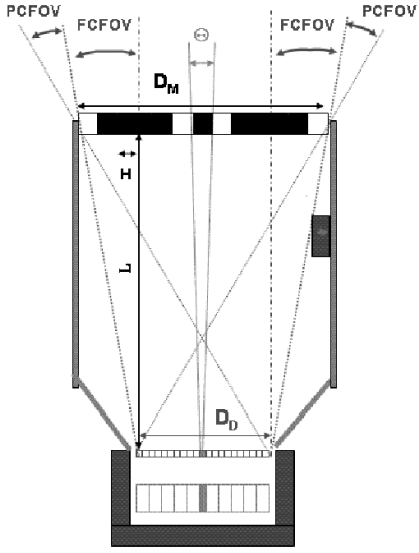, width=60mm}}
\caption{
Scheme (not in scale) of the IBIS imaging system:
a coded mask of $11.2 \times 11.2 \times 16$ mm$^3$ tungsten elements,
an upper low-energy detector plane (red) of 128$\times$128 CdTe crystals
($4 \times 4 \times 2$ mm$^3$, pitch pixel-to-pixel = 4.6 mm),
a high energy bottom detector plane (green) of 64$\times$64 CsI bars 
($8.4 \times 8.4 \times 30$ mm$^3$, pitch pixel-to-pixel = 9.2 mm),
an active BGO veto system (blue) composed of 19 crystals viewed by photomultipliers,
a passive shield made by a tube and a Hopper shield (grey),
a calibration unit attached to the tube.
The Fully Coded (= ${\rm arctan}[(D_{\rm M}-D_{\rm D})/L]$), and the Partially Coded, 
(= ${\rm arctan}[(D_{\rm M}+D_{\rm D})/L]$), Field of Views are 
indicated along with the telescope angular resolution (FWHM) $\Theta = {\rm arctan}(H/L)$,
where $D_{\rm D}$ and $D_{\rm M}$ are the detector and mask dimensions, 
$H$ is the width of the mask elements and $L$ 
the distance mask-detector ($\approx$~3200~mm for ISGRI).}
\label{fig:Ibis}
\end{figure}
\begin{figure}
\centering
\centerline{\epsfig{figure=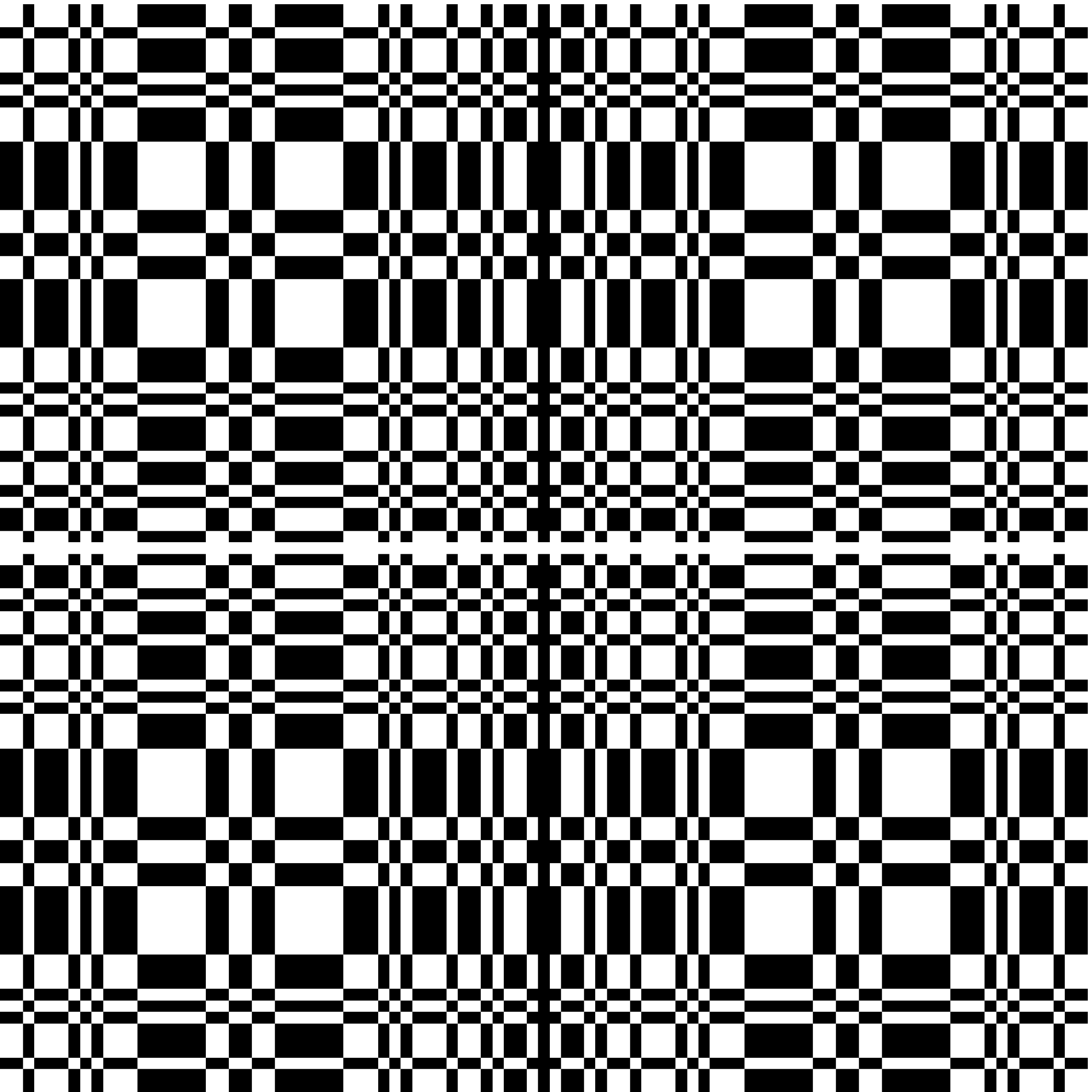,width=38mm}
\epsfig{figure=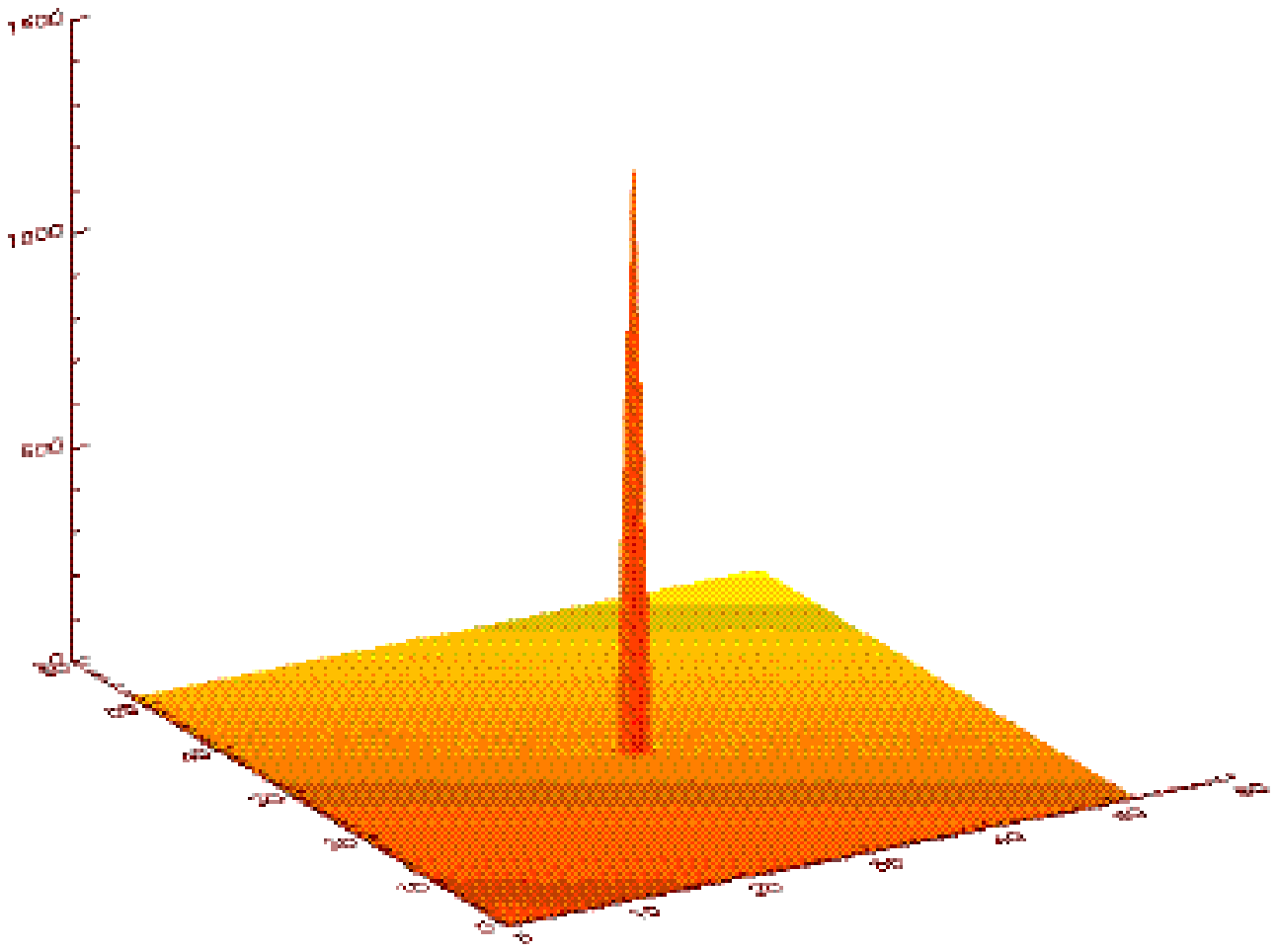,width=45mm}
}
\caption{The IBIS mask pattern of 95$\times$95 elements (left) is formed 
by a replicated 53$\times$53 MURA basic pattern, whose cyclic autocorrelation 
(right) is a $\delta$ function.}
\label{fig:Mura}
\end{figure}

The IBIS coded mask imaging system (Fig.~\ref{fig:Ibis}) includes a replicated 
Modified URA (MURA)(Gottesman $\&$ Fenimore 1989) mask of tungsten elements (Fig.~\ref{fig:Mura}) 
and 2 pixellated gamma-ray detector planes, both approximately of the same size 
of the mask basic pattern: ISGRI, the low energy band (15 keV - 1 MeV) camera
(Lebrun et al. 2003) and PICsIT, sensitive to photons between 175~keV and 10~MeV, 
disposed about 10~cm below ISGRI (Di Cocco et al. 2003).
The physical characteristics of the IBIS telescope define a FCFOV of
8$^{\circ}$~$\times$~8$^{\circ}$ and a total FOV (FCFOV+PCFOV) of 
19$^{\circ}$~$\times$~19$^{\circ}$ at half sensitivity and of
29$^{\circ}$~$\times$~29$^{\circ}$ at zero sensitivity.
The nominal angular resolution (FWHM) is of 12$'$. ISGRI images  
are sampled in 5$'$ pixels while PICsIT images in 10$'$ pixels.
The MURAs are nearly-optimum masks and a {\it correlation inverse}
is obtained by setting $G = 2 M - 1$ (i.e. $G=+1$ for $M=1$, $G=-1$ for $M=0$) apart from
the central element which is set to 0. Simple correlation between such array and the
detector plane array provides a sky image of the FCFOV where point sources appear as spikes
of approximately the size (FWHM) of 1 projected mask element (12$'$) with flat sidelobes (Fig. 2).
The detailed shape of the System Point Spread Sunction (SPSF), the final response 
of the imaging system including the decoding process to the a point source, 
actually depends also on the instrument features 
(pixel size, detector deadzones, mask thickness, etc.) and on the decoding process 
(Gros et al. 2003).
For this simple reconstruction (sum of transparent elements and subtraction of opaque ones)
and assuming Poissonian noise,
the variance in each reconstructed sky pixel of the FCFOV is constant 
and simply given by 
$V = G^2 \star D = \sum_{kl} D_{kl} $, i.e. the total counts recorded by the detector.
Therefore the source signal to noise is simply
$$S/N = {C_{\rm S} \over { \sqrt{C_{\rm S} + C_{\rm B}} } } = 
{{\rm Reconstructed~Source~Counts} \over \sqrt{~\rm Total~Counts}} $$
Sources outside the FCFOV but within the PCFOV still project part of the mask on the detector
and their contribution can (and must) be reconstructed by properly extending the analysis
in this region of the sky. 
In the PCFOV the SPSF does have secondary lobes (Fig.~\ref{fig:MuraPsf}),
the sensitivity decreases and the relative variance increases towards the edge of the field.
In this part of the field also FCFOV sources will produce side lobes (see Sect.~2).

\section{Sky image deconvolution}
\begin{figure}
\centering
\centerline{\epsfig{figure=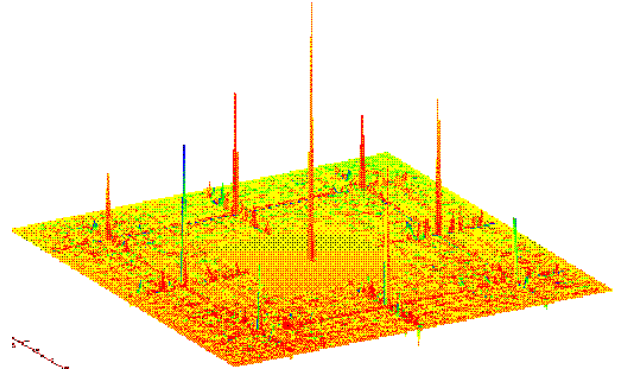 ,width=80mm}}
\caption{The System Point Spread Function 
over the complete FOV for the IBIS/ISGRI telescope.} 
\label{fig:MuraPsf}
\end{figure}

The discrete deconvolution in FCFOV can be extended to the total
(FC+PC) FOV by performing 
the correlation of the detector array $D$ in a non cyclic form with 
the $G$ array extended and padded with 0 elements outside the mask.
Since the number of correlated (transparent and opaque) elements in the PCFOV 
is not constant as in the FCFOV, 
the sums and subtractions for each sky position must be balanced and 
renormalized.
This can be written by 
$$ 
S_{ij}=  { \sum^{ }_{kl}G^+_{i+k,j+l}W_{kl}D_{kl} \over  \sum^{ }_ {kl}G^+_{i+k,j+l}W_{kl}} 
- {\sum^{ }_{kl}G^-_{i+k,j+l}W_{kl}D_{kl} \over \sum^{ }_{kl}G^-_{i+k,j+l}W_{kl}}
$$
where the decoding arrays are obtained from the G array by setting
$G^+ = G$~for~$G > 0$, $G^+ = 0$~for~$G \le 0$ and 
$G^- = -G$~for~$G < 0$, $G^- = 0$~for~$G \ge 0$, 
and then are enlarged and padded with 0's outside the mask region. The sum is performed 
over all detector elements $k,l$ and $i,j$ run over all sky pixels (both of FC and PCFOV).
In the FCFOV we obtain the same result of the standard FCFOV correlation.
To consider effects such as satellite drift corrections (see Goldwurm 1995), 
dead areas, noisy pixels or other instrumental effects which require different weighing 
of pixels values, a specific array $W$ is used. 
For example $W$ is set to 0 for noisy pixels or detector dead areas and corresponding 
detector pixels are not included in the computation.
The variance, which is not constant outside the FCFOV, is computed accordingly by
$$ 
V_{ij}=  { \sum^{ }_ {kl}(G^+_{i+k,j+l}W_{kl})^2D_{kl} \over  (\sum^{ }_ {kl}G^+_{i+k,j+l}W_{kl})^2} 
+ {\sum^{ }_{kl}(G^-_{i+k,j+l}W_{kl})^2D_{kl} \over (\sum^{ }_ {kl}G^-_{i+k,j+l}W_{kl})^2} 
$$
since the cross-terms $G^+$ and $G^-$ vanish. 
In the FCFOV the variance is approximately constant
and equal to the total number of counts on the detector.
The values computed by these formula are then renormalized
to obtain the images of reconstructed counts per second from the whole detector
for an on-axis source.
Since mask elements are larger than detector pixels, 
the decoding arrays are derived as above 
after $G$ has been sampled in detector pixels by projection and redistribution of 
its values on a pixel grid. 
For non integer sampling of elements in pixels
(the IBIS case) $G$ will assume continuous values from -1 to 1. 
Different options can be adopted for the fine sampling of the decoding array 
(Fenimore $\&$ Cannon 1981, Goldwurm 1995), the
one we follow optimizes the point source signal to noise 
(while slightly spreading the SPSF peak).
The SPSF, for an on-axis source and the IBIS/ISGRI configuration,
obtained with the described deconvolution is shown in Fig.~\ref{fig:MuraPsf}. 
Note the central peak and the flat level in the FCFOV, 
and the secondary lobes with the 8 main ghosts of the source peak
at distances multiple of the mask basic pattern in the PCFOV.
This procedure can be carried out with a fast algorithm by
reducing previous formulae to a set of correlations computed by means of FFTs. 

\section {Non-uniformity and background correction}

In most cases and in particular at high energies the background is not 
spatially uniform.
Further modulation is introduced on the background and source terms
by the intrinsic detector non-uniformity.
The modulation is magnified by the decoding process (the term $G \star B$ in 
Sect. 1) and strong systematic noise is generated in the deconvolved images 
(Laudet $\&$ Roques 1988).
The resulting systematic image structures have spatial frequencies 
similar to those of the modulation on the detector plane.
For the IBIS detector, intrinsic non-uniformity is generated by 
variations in efficiency of single detectors or associated electronics.
Anticoincidence effects and scattering may also induce features at
small spatial scales, of the order of the mask element size (Bird et al. 2003).
The modulation induced by the external background is normally 
a large scale structure dominated by the differences in efficiency of
the veto system and of  multiple event tagging.
The background intensity may be variable on timescales from hours/days
(activation in radiation belts, solar flares)
to weeks/months (orbit circularization, solar cycle modulation)
(Terrier et al. 2003).

For the PICsIT layer, two more effects should be taken into account: the
modulation at the border of the detector semimodules (Di Cocco et al. 2003) 
caused by the onboard multiple
reconstruction system, and the cosmic-rays induced events. The
former is due to the fact that multiple events are reconstructed by
analysing events per semimodule. When a multiple event is detected by two
semimodules, it is considered as two single events. Since this occurs at
the borders of the semimodules, there is an excess of
single events in these zones, and a corresponding deficit of
multiple counts in the same pixels. Obviously this effect becomes
important as the energy increases. In Fig. 5 it is shown an example of a single
event shadowgram not yet corrected, where the border effects are clearly
visible, and after the correction for background and non-uniformities
(Natalucci et al. 2003). The second effect is due to
spurious events caused by cosmic-rays interaction with CsI crystals 
(Segreto et al. 2003). This is relevant only at low energies ($<$ 250 keV) and
results in a loss of sensitivity of about a factor of 4. It is possible to
remove these fake events only if PICsIT is in photon-by-photon mode, while
on the histograms it is possible to apply only an {\it a-posteriori correction},
by rescaling those pixels that, after the background subtraction, appear
to be still noisy.

If the shape of the background does not vary rapidly, regular observations of 
source empty fields can provide measures of the background spatial distribution
which can be used to flat-field the detector images prior to the decoding
(contribution of weak sources is smoothed out by summing images
corresponding to different pointing aspects) (Bouchet et al. 2001).
If $U$ is the detector non-uniformity and $B$ the background structure, 
then the recorded detector image during an observation is
$D=(S \ast M) U + B$ and a basic correction can be performed by 
$ (D - b B_{EF}) / U_C  $, where $B_{\rm EF}$ is obtained from the empty field observations,
$U_{\rm C}$ is an estimate of the detector non-uniformity from ground calibrations or 
Monte Carlo modelling and b is a normalization factor.
If $U_C$ is a good estimate of $U$ and $B_{\rm EF}$ is close to $B$ we
obtain the needed correction. The normalization $b$ is estimated either using the 
relative exposure times or the total number of counts in the images.
\begin{figure}
\centering
{\epsfig{figure=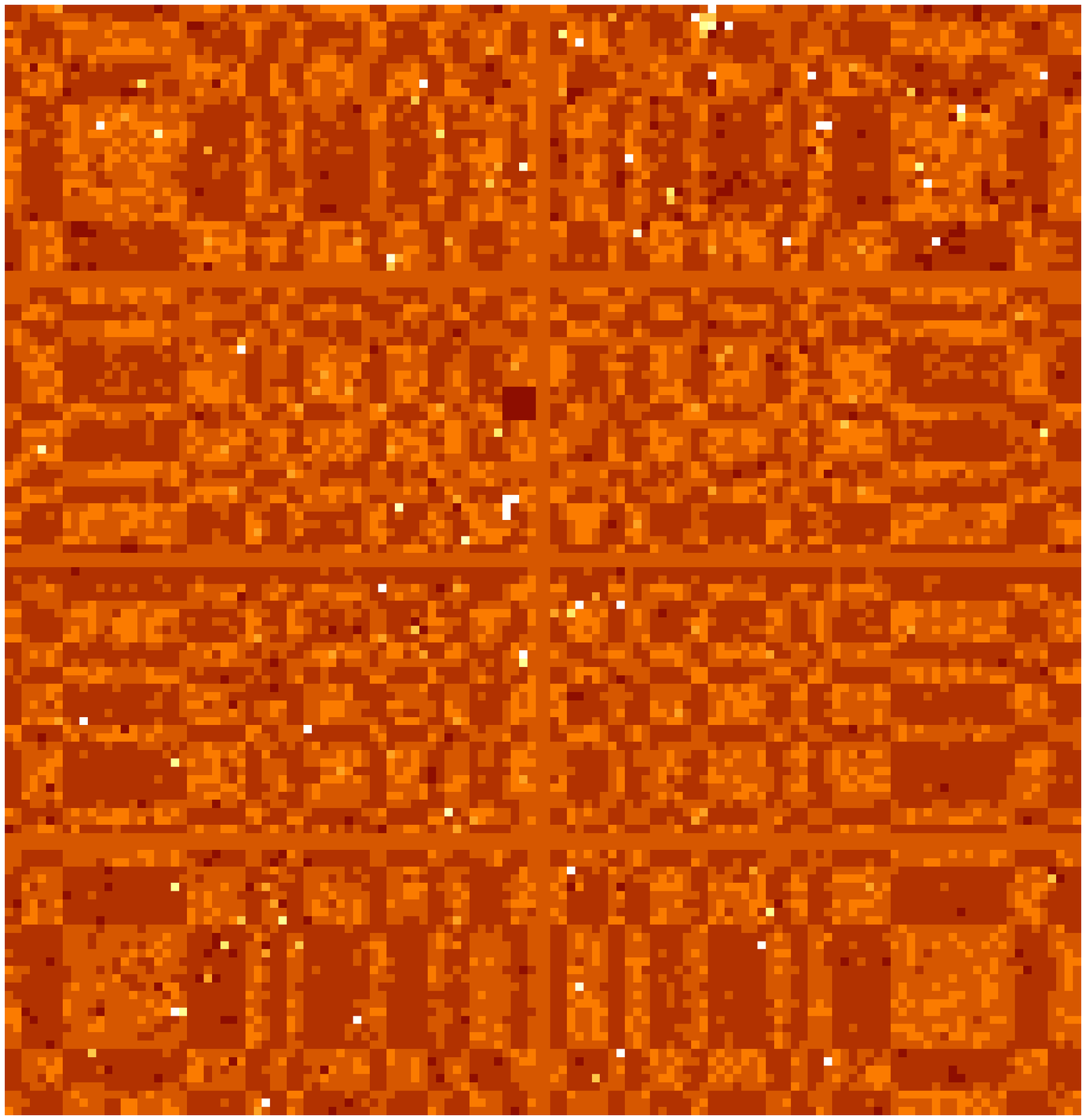,width=42mm} 
\epsfig{figure=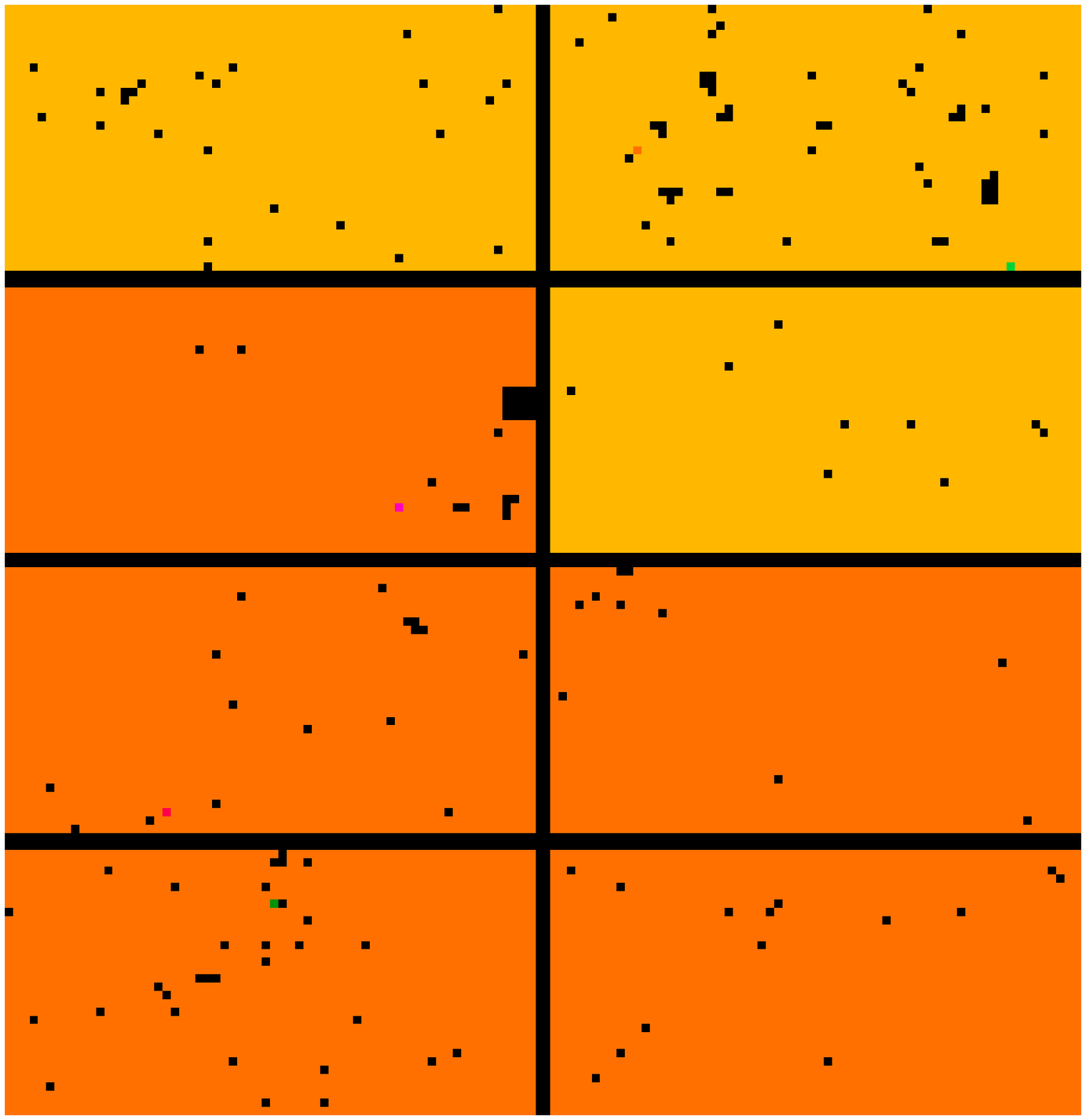,width=42mm}
\epsfig{figure=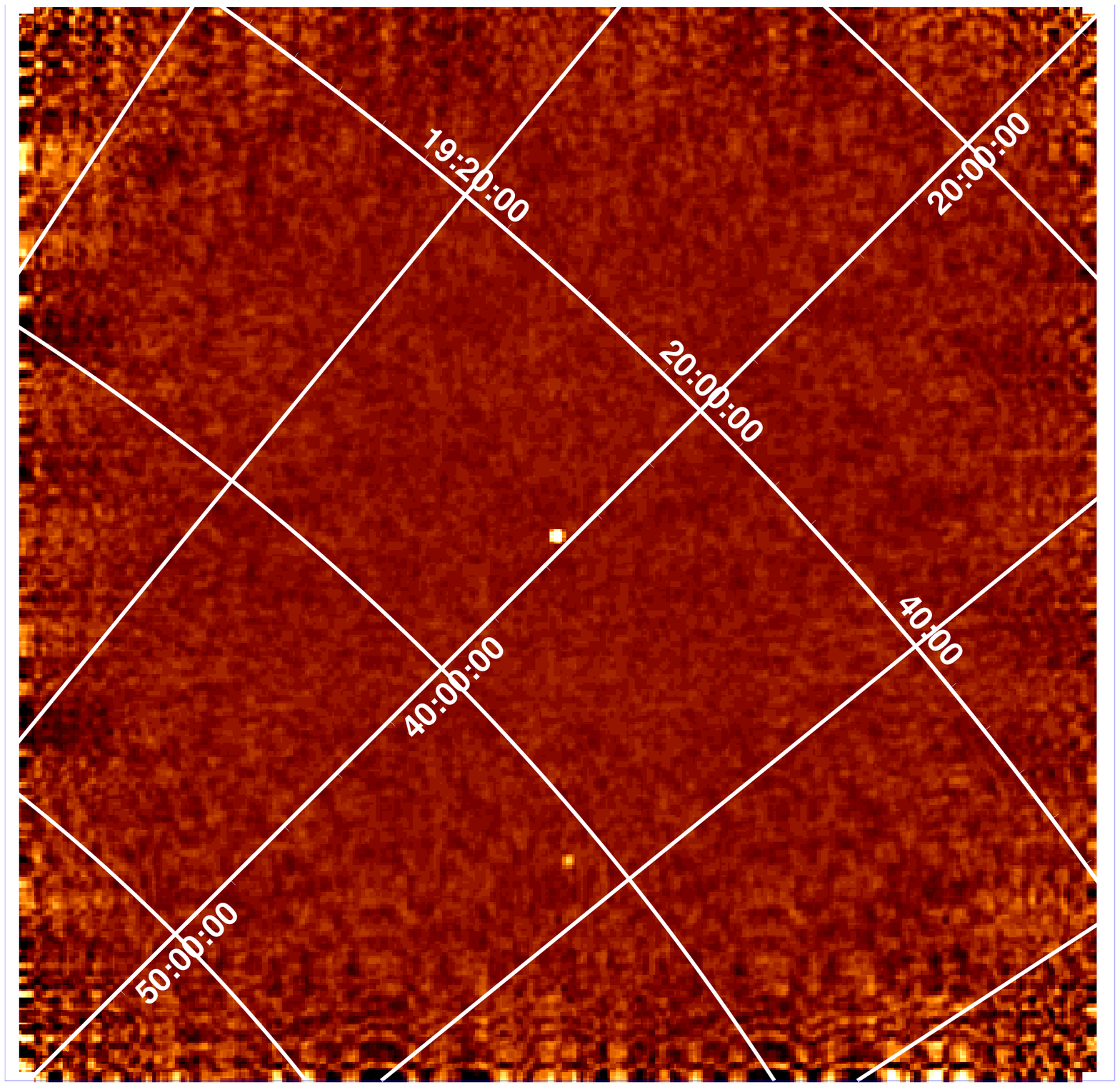,width=42mm}
\epsfig{figure=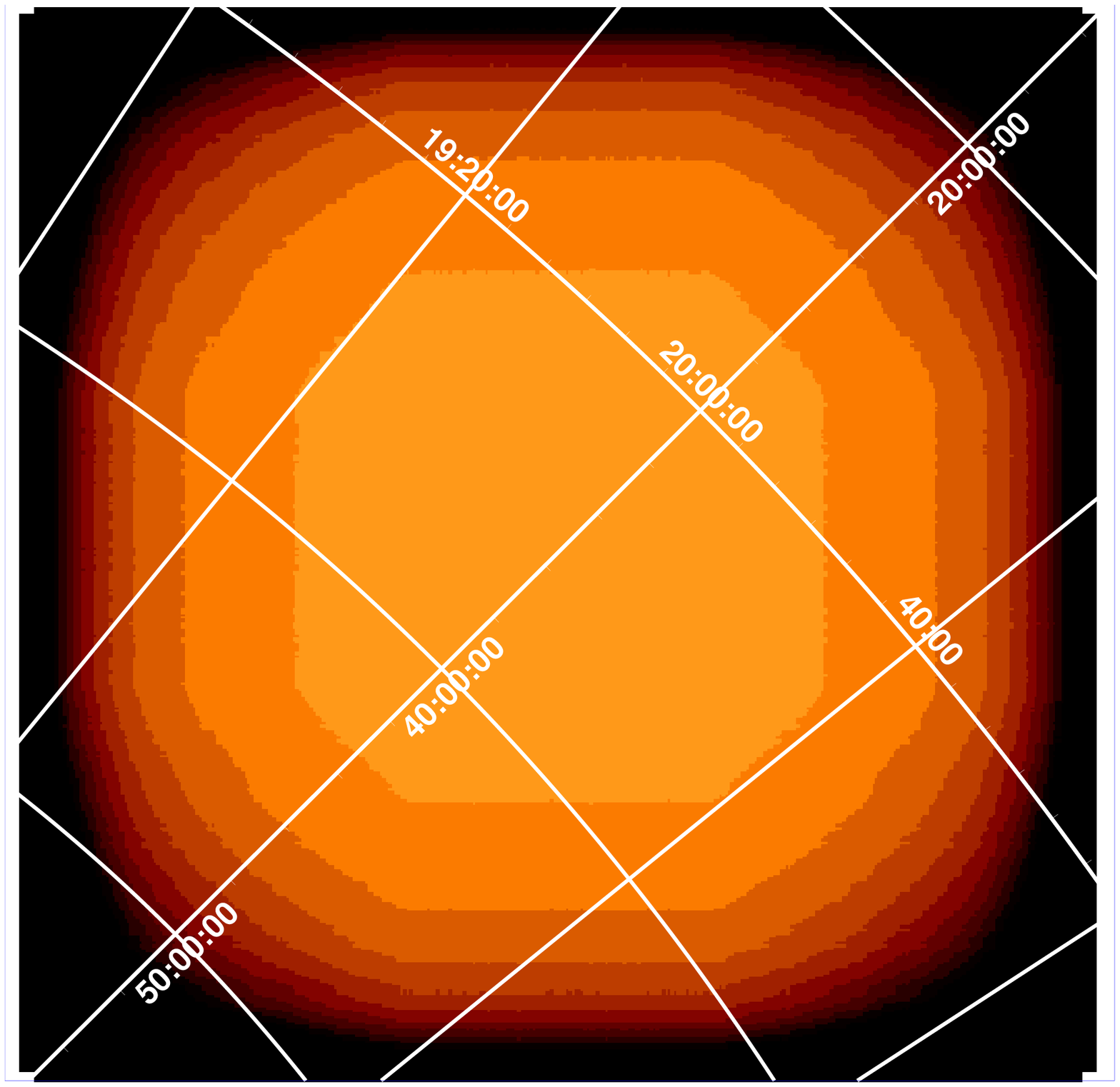,width=42mm}
}
\caption{Results of reconstruction procedure applied to IBIS/ISGRI 
data of the Cygnus region: 
detector image after background correction (up, left), 
associated efficiency image (up, right), 
decoded and cleaned sky image in the whole FOV (FC+PC) (down, left), 
associated variance (down, right). The algorithm detected, and identified 2 sources: 
Cyg X-1 at the center of the FOV and Cyg X-3 in the PCFOV.}
\label{fig:CorVar}
\end{figure}

\section{The IBIS scientific data analysis}

The specific procedure for the IBIS data analysis starts from the data files
obtained with the Integral Science Data Center (ISDC) (Courvoisier et al. 2003) 
preprocessing and performs a number of analysis steps 
which are hereafter described for the IBIS {\it standard and photon-photon modes}
and the ISGRI and PICsIT data only.

The ISDC preprocessing decodes the telemetry packets, prepares the scientific
and housekeping (HK) data with the proper satellite onboard times. 
This analysis task also 
computes the good time intervals by checking the satellite attitude, 
the telemetry gaps and the instrument modes and by monitoring the technological parameters. 
For the IBIS standard mode (Ubertini et al. 2003, Di Cocco et al. 2003) 
the following data sets are provided, for each elementary 
observation interval (science window) which corresponds to a constant pointing 
or slew satellite mode: \\
- ISGRI event list: 
position, arrival time, pulse-height channel, rise-time. \\
- PICsIT spectral-image histograms: 
2 sets of $64 \times 64$ pixel images in 256 energy channels, one for the
integrated single events and the other for the integrated multiple events. \\
- PICsIT spectral timing histograms: count rates from the whole PICsIT camera 
integrated in a short time interval (0.97-500 ms, default: 2 ms) for up to 
8 energy bands (default: 4 bands). \\
- Compton events list: 
ISGRI position, PICsIT position, ISGRI deposited energy, 
PICsIT deposited energy, ISGRI risetime, arrival time of events in coincidence between 
ISGRI and PICsIT.\\
For the IBIS photon-photon mode the same ISGRI and Compton data are provided 
while PICsIT histograms are replaced by \\
- PICsIT event list for single and double events: position, deposited energy, arrival time.
\\
The first tasks performed by the IBIS specific software are the computation of
the livetimes of the single pixels, of the deadtimes and of the 
deposited energies for all events. Detector images are then built and corrected.
Sky images are reconstructed for each pointing and then combined to
obtain mosaic of sky images of the whole observation.
Source spectra are extracted by performing image binning and correction 
on small energy bands and then comparing resulting images to source models.
All data products are written in files with the standard FITS and OGIP format
in order to allow the use of standard high-level data analysis packages 
like FTOOLS, XSPEC and DS9.

\subsection{Pixels livetimes, deadtimes and energy corrections}

From the data of each single science window, prepared by the standard ISDC preprocessing,
selected IBIS housekeeping (HK) data are first analyzed to prepare correction parameters
for the scientific analysis. The HK tables reporting the initial status (on/off) of detector pixels, 
their low energy thresholds and their status evolution during the 
observations are decoded and the information organized to allow computation of livetime of
each single pixel during the science window. 
The deadtime and its evolution is computed from the countrates 
reported in the HKs for the different IBIS datatypes, also including the time loss
of random coincidences during veto rejection or calibration source tagging.

The ISGRI event pulse height channel and risetime information in the science data list 
are then used to reconstruct event deposited energy (in keV) by correcting the pulse height amplitude
for the charge loss effect and for gains and offsets (Lebrun et al. 2003).  The correction is
performed using look-up tables (LUT) derived from ground and inflight calibrations (Terrier et al. 2003).
PICsIT energies are reconstructed on board using gain and offset LUT
before histograms are accumulated. 
A further correction on the PICsIT single events can
be performed on ground by using pixel gain and offsets
values refined according to temperature variations (Malaguti et al. 2003).

\subsection{Image binning and uniformity-background correction}
For each pointing science window the event list or the histograms 
are then binned in detector images for the specified energy bands and for
a given risetime interval.
The procedure computes an efficiency (or exposure) map combining informations from
good time intervals, telemetry or data gaps, deadtimes, livetime of each pixel, 
and low energy thresholds.
The map gives an exposure of each pixels considering all these effects
and relative to an effective exposure time of the pointing.

ISGRI images are then enlarged to 130~$\times$~134 pixels and PICsIT images 
to 65~$\times$~67 pixels to include the deadzones between the modules.
Residual noisy pixels, i.e. pixels with 
large values ($> 5-10 ~\sigma$) above the average are then identified and 
efficiencies of noisy and deadzone pixels are set to 0.
The detector image pixels are divided by the non-zero-values of the efficiencies 
to renormalize their values. 
For efficiencies equal to 0 the weighting array $W$ is set to 0 and the pixels are not 
included in the image or spectral reconstructions.
The correction for detector and background non-uniformity is 
then applied using the background images derived from empty field observations and 
the detector uniformity maps (Terrier et al. 2003, Natalucci et al 2003).
For the spectral extraction however the correction is not perfomed at this level,
background and non-uniformity are accounted for directly in the fitting procedure.
\begin{figure}
\centering
{\epsfig{figure=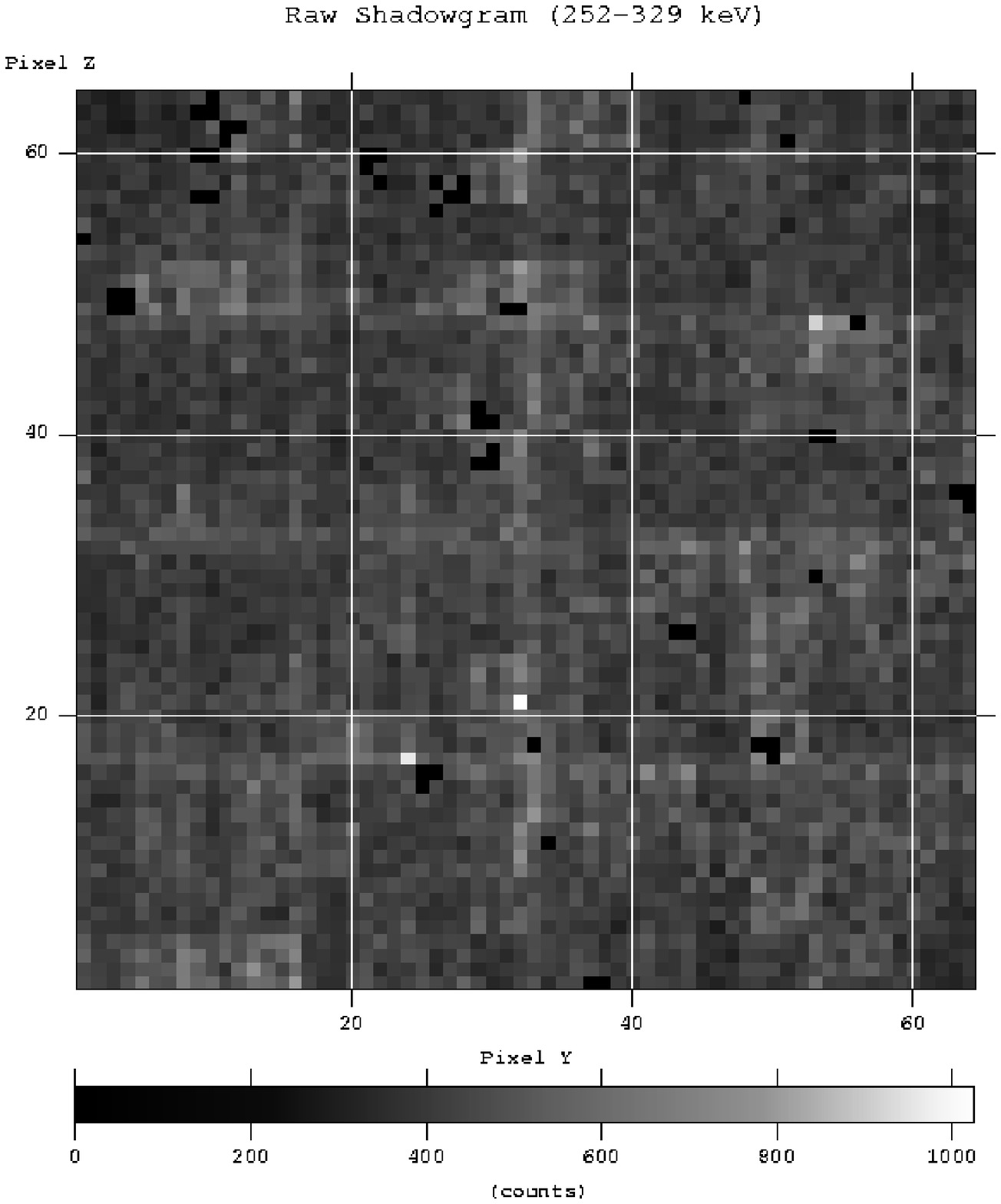 ,width=42mm} 
\epsfig{figure=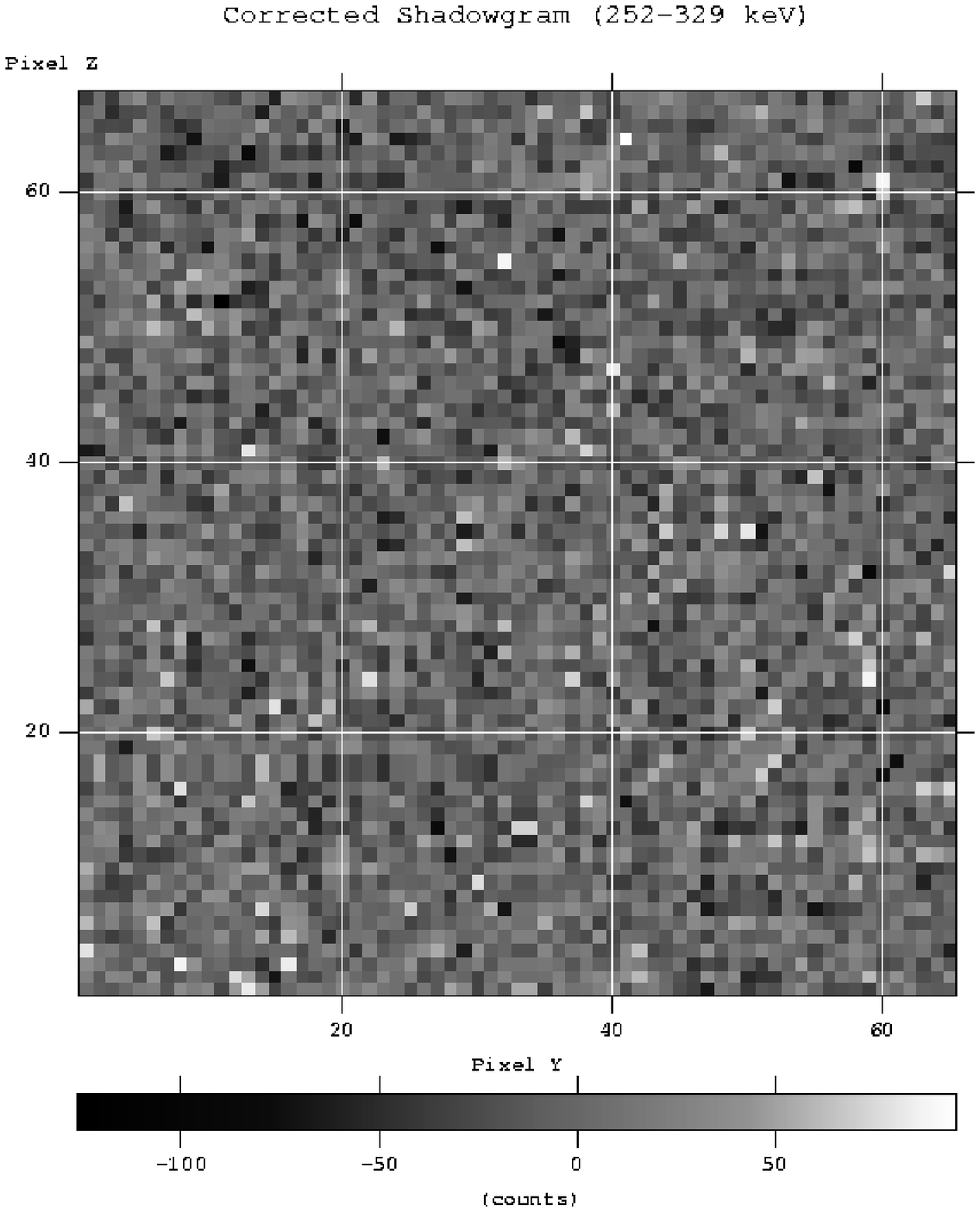,width=42mm}
}
\caption{Results of the background correction on a PICsIT single event
shadowgram (252-329 keV) of the Crab region. 
The raw image (left) shows presence of dead pixels and  count excess on the module
borders, not visible in the background corrected image (right).}
\label{fig:PICubc}
\end{figure}

\subsection{Iterative sky image reconstruction and cleaning}
From the corrected detector images the analysis procedure decodes the 
sky images using the algorithm described in Sect.~2 and iteratively searches for significant peaks
in the image. The signal to noise levels (S/N) of detection can be set by the users, and the 
search can be preferencially performed or even totally restricted to the sources included 
in an input catalogue. 
The search procedure is a delicate process due to the presence of the 8 ghosts (per source)
and is sensitive to the choices of the S/N levels of search 
in particular if the background is not fully corrected.
The first significant peak is detected and fitted with an analytical approximation 
of the SPSF to finely determine the position of the source and its flux (Gros et al. 2003).
The model of the shadowgram projected by the source at the derived 
position is computed, decoded, normalized to the observed excess
and subtracted from the image. 
Such a source response model must take into account the
mask element thickness, the effects at the mask border, the
different opacity of the mask support and of the passive shield, and
the transparent zones in or between these elements. However the model includes only 
the geometrical absorption effects. 
Proper account of scattering and energy redistribution is included in the energy response.
The same routine which computes the model of source shadowgram on the detector
is used in the spectral extraction.
The source parameters (position, flux, and signal to noise) 
are stored in an output file and the procedure uses the source catalogue
to identify the source. A second excess is then searched and cleaned and the 
procedure continues till all sources or excesses are identified and cleaned.
The image of the main peak of each source is finally restored in the 
sky image while the secondary lobes are excluded.
Units of intensity images are in reconstructed counts per second
for an on-axis source, and a variance image is computed in the proper units. 

\subsection{Image mosaic and final image analysis}
The reconstructed and cleaned sky images of each pointing science window 
are then rotated, projected over a reference image 
and summed after a properly weighting with their variance and exposure time.
There are two ways to implement the rotation of an image.
Either the pixel value is redistributed around the reference pixels on which is projected
or its value is entirely attributed to the pixel whose center is closer to the
projected center of the input pixel. 
In the first case the source is slightly smoothed and enlarged but the source centroid
is well reconstructed,
while in the second case discrete effects can produce a bias in the source centroid 
but the intensity will be better reconstructed and the source less spread.

After the mosaic step is completed the routine restart the search and analysis of 
point sources in the mosaicked sky image, again using a catalogue for the 
identification.
Results of fluxes and positions are reported in output. Since the point source location 
error is inversely proportional to the source signal to noise (Gros et al. 2003), 
location of weak sources is better carried out on mosaicked images.
Note that the search for significant excesses (both for single pointing 
and for the mosaic)
must be performed taking into account that these are {\it correlation images} and 
considering the number of independent trials that are made by testing all pixels of the
reconstructed sky image. 
The critical level at which an unknown excess is significant must be 
increased from the standard 3~$\sigma$ value to typically 5~-~6~$\sigma$ 
(see e.g. Caroli et al. 1987). In the case of residual (background) systematic modulation, 
the detection level must be further increased (for example by a factor given by the ratio
of measured to computed image standard deviation).
\begin{figure}
\centering
{\epsfig{figure=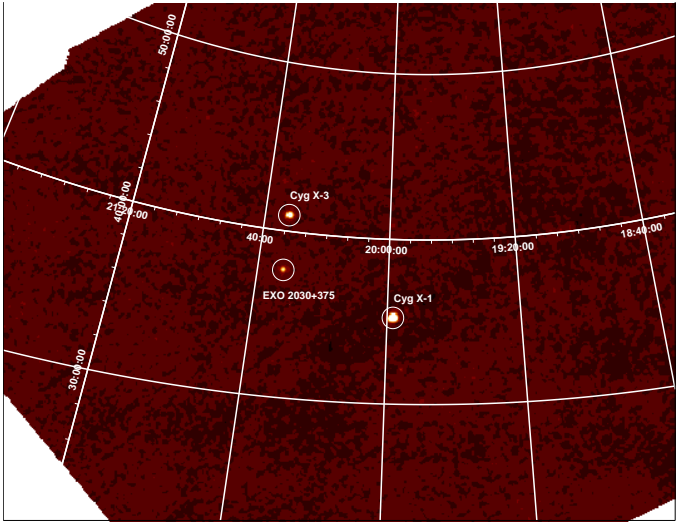 ,width=90mm} 
}
\caption{Sky image mosaic of the Cygnus region in the 15-40 keV band
composed with IBIS/ISGRI data from 130 pointings for a total exposure of
145 ks. Three sources
are detected at high significance level: Cyg X-1, Cyg X-3 and EXO 2030+375.}
\label{fig:CygMos}
\end{figure}

\subsection{Source spectra and light curve extractions}
Once the positions of the active sources of the field are known, their fluxes
and count spectra are extracted for each pointing science window
in predefined energy bins.
Detector images and their efficiencies are built for these energy bins,
enlarged and then searched for residual noisy pixels.
For each source and each energy band a shadowgram model is then computed using the
detailed modelling algorithm described above. 
If $P_{kl}^n (\Delta E_h)$ is the model for the source n and the energy bin $\Delta E_h$,
the total sky plus background model $T$ at the pixel $k,l$ is defined as
$$ T_{kl}^m (\Delta E_h)= \sum_{n = 1}^{m}  f_n^h  P_{kl}^n(\Delta E_h) E_{kl}^h U_C^h + b^h B_{kl}(\Delta E_h) E_{kl}^h$$
where the sum is extended to all "active" sources of the region and 
where the background image $B$ is derived from the empty field data, the $U_{\rm C}$ is the estimated
detector non-uniformity and $E$ is the efficiency image.
This model is fitted to the enlarged and noisy-pixel-corrected detector images $D_{k,l}$  
by using the maximum likelihood technique with Poissonian distribution.
The optimization gives the multiplicative factors $f_n^h$ and $b^h$ for each energy band
which (after proper normalizations and division by the exposure time), provide  
the source and background count rate spectra.
It is important to perform the simultanous fit for the models of all the active sources 
of the field to avoid contamination by other sources of the interested spectrum.
Spectra of the same source collected in different pointing science windows can then be summed
to obtain an average spectrum during an observation.  The spectra are compared 
to physical spectral models convolved with the energy response function  
of the instrument (Laurent et al. 2003) in order to evaluate source physical parameters.
Finally, a similar procedure, 
where the images are binned in large energy bands and small time intervals,
provides source light curves on time bins shorter than science
windows for the predefined active sources in the field of view. 
For PICsIT operating in standard mode, light curves of the count rates registered 
by the whole detector can be generated directly from the spectral timing data and used to
search for characteristic variability signatures like pulsations or bursts.

\section{Final remarks}

\begin{figure}
\centering
{\epsfig{figure=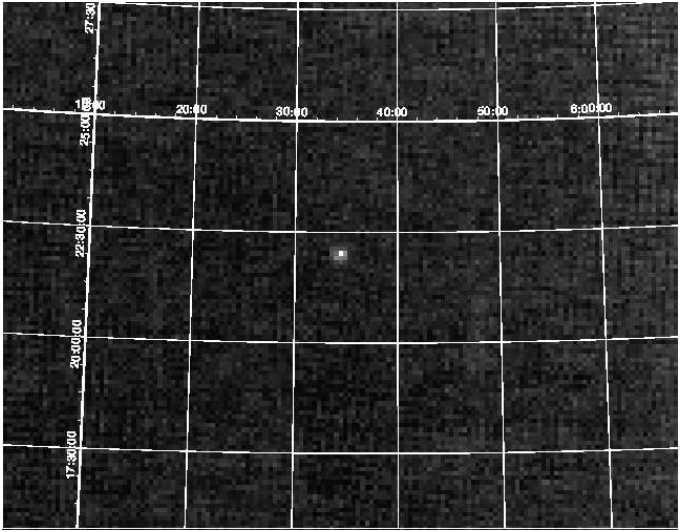,width=90mm}
}
\caption{IBIS/PICsIT reconstructed image of the Crab in the energy band 252-329 
keV obtained from the IBIS scientific pipeline applied to 55 science windows 
for a total of 132~ks of exposure.}
\label{fig:CraPic}
\end{figure}

\begin{figure}
\centering
{\epsfig{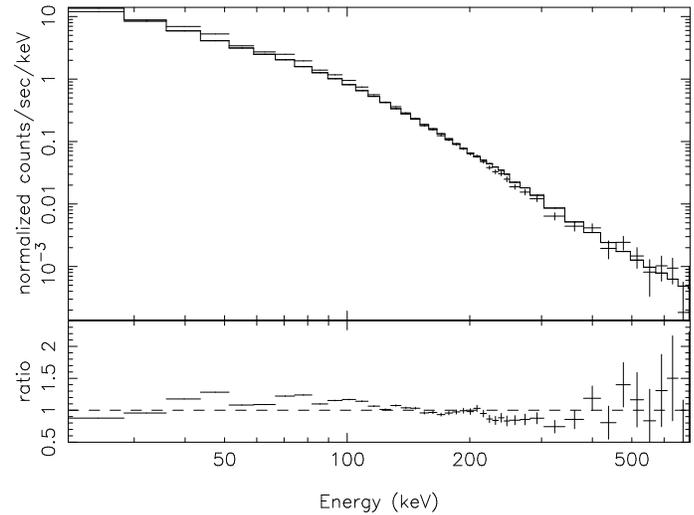} 
}
\caption{IBIS/ISGRI reconstructed count spectrum of the Crab 
from data of a 90 ks exposure pointed observation. 
The best fit model of a power law and ratios
are also shown. The ratios show presence of 10$\%$ systematic 
erros in particular in the 20-100 keV range.}
\label{fig:CraSpe}
\end{figure}

The software has been tested using inflight data in several conditions 
and the performance is satisfactory considering the early phase of the mission.
Background corrections have been already implemented for the analysis of PICsIT
data while not yet fully used for the ISGRI analysis (but see Terrier et al. 2003).
Fig.~\ref{fig:PICubc} shows the correction operated on the Crab nebula PICsIT images
as described in Sect. 3.
In Fig.~\ref{fig:CygMos} the resulting sky image mosaic obtained from the IBIS scientific pipeline 
applied to about 130 elementary pointings 
on Cygnus X-1, illustrate the quality of the scientific analysis 
presently available. 
Images are well reconstructed and cleaned and sources well detected and finely positioned, 
with typical absolute error radii (90$\%$ confidence level) of less 1$'$ 
for strong source (S/N~$>$~30-40)
and less than 3$'$ for moderate or weak ones (S/N~$\approx$~10-40) (Gros et al. 2003).
Fig.~\ref{fig:CraPic} shows a reconstructed PICsIT image of the Crab, a very bright
constant and pointlike source for IBIS, often used as calibration source for 
high energy instruments.
Fig.~\ref{fig:CraSpe} reports a reconstructed Crab count spectrum 
obtained with the described scientific analysis procedures from the ISGRI
data of an observation pointed on the source and performed in {\it staring mode}
(constant attitude asset during the observation, see Winkler et al. 2003). 
The spectrum is compared to a model of a simple power law
using the instrument energy response files.
The best fit in the range 20-700 keV is found for a slope of 2.05 which is 
compatible with the high energy spectrum of the Crab. 
To have acceptable $\chi ^2$, systematic errors of about 10$\%$ have 
to be included. 

These examples give an idea of the performance of the analysis s/w and of
the calibration and response files presently available.
The procedures described here have been integrated in the 
ISDC analysis system and constitute the core of the IBIS data analysis software of the
Quick Look and Standard Analysis run at the ISDC and of the 
INTEGRAL Data Analysis System package.
The IBIS analysis software and associate calibration files are constantly improved
and integrated in the following versions of the ISDC system.
The present set has already been used to derive a number of interesting 
results from the IBIS data 
of the high energy sources observed with INTEGRAL (see this volume).


\section*{Acknowledgments}
A.Gros and P.D. acknowledge financial support from the French Spatial Agency (CNES).
L.F. acknowledges financial support from the Italian Spatial Agency (ASI) and the
hospitality of the ISDC.

\end{document}